# High temperature superconductivity in sulfur hydride under ultrahigh pressure: A complex superconducting phase beyond conventional BCS

Annette Bussmann-Holder[1], Jürgen Köhler[1], M.-H. Whangbo[2],
Antonio Bianconi[3,4] and Arndt Simon[1]

[1]Max-Planck-Institute for Solid State Research, Heisenbergstr. 1, D70569- Stuttgart, Germany
[2]Department of Chemistry, North Carolina State University, Raleigh, NC 27695-8204, USA
[3]RICMASS, Rome International Center for Materials Science Superstripes, Via dei Sabelli 119A, 00185 Rome, Italy
[4]MEPhI, Moscow Engineering Physics Institute, National Research Nuclear University, Solid State and Nanosystems Physics, Kashirskoye sh. 31, Moscow 115409, Russia

**key words**: multi-gap superconductivity, multiple-condensates, steep and flat bands, high temperature superconductivity, pressurized sulfur hydrides.

*Abstract*

The recent report of superconductivity under high pressure at the record transition temperature of $T_c$=203K in pressurized $H_2S$ has been identified as conventional in view of the observation of an isotope effect upon deuteration. Here it is demonstrated that conventional theories of superconductivity in the sense of BCS or Eliashberg formalisms cannot account for the pressure dependence of the isotope coefficient. The only way out of the dilemma is a multi-band approach of superconductivity where already small interband coupling suffices to achieve the high values of $T_c$ together with the anomalous pressure dependent isotope coefficient. In addition, it is shown that anharmonicity of the hydrogen bonds vanishes under pressure whereas anharmonic phonon modes related to sulfur are still active.

## 1. Introduction

Early predictions of high temperature superconductivity in hydrogen bonded systems are related to the idea that the light H atom favors high frequency vibrations thus supporting high temperature superconductivity [1-4]. Correspondingly there were many attempts to solidify hydrogen containing materials, to make them metallic and superconducting under pressure [5-7]. A major focus was laid on prediction of hydrogen sulfide structural phase transition at high pressure [8,9] and experiments under high pressure reported the appearance of different structural modifications with increasing pressure including the insulator to metal transition. However, superconductivity was never observed until recently when a pressure larger than 110 GPa [10] was applied to $H_2S$ [10]. The report of an isotope effect on $T_c$ was released with the isotope coefficient $\alpha$ =0.3 at 200 GPa [10] smaller but close to the BCS value $\alpha$ =0.5.

A. Bussmann-Holder, J. Köhler, M.-H. Whangbo, A. Bianconi, A. Simon.

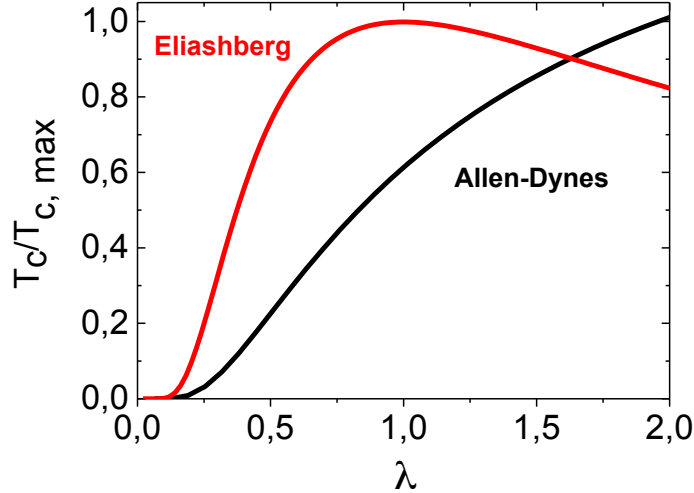

**Figure 1.** The typical behaviors of the normalized superconducting transition temperature $T_c/T_{c,max}$ as a function of the electron-phonon coupling parameter $\lambda$ in *a single band isotropic* BCS model. While within the Allen-Dynes scheme (black line) the critical temperature increases with the electron-phonon coupling parameter $\lambda$, in the Eliashberg theory (red line) the critical temperature reaches a maximum and then decreases by increasing further the electron-phonon coupling parameter. The maximum $T_{c,max}$ depends on the phonon frequency. Note, that the maximum coupling $\lambda = 2$ used here is already far beyond strong coupling theory.

In view of the apparent absence of any kind of magnetism, pressurized $H_2S$ was classified as a material where pairing is mediated by phonons and therefore called a "conventional" superconductor [10-15], in contrast to cuprates and pnictides called "unconventional" superconductors. In these unconventional superconductors, superconductivity emerges by doping a magnetic phase, competes and coexists with intertwined spin fluctuations. However these unconventional superconductors show a complex landscape due to quenched disorder and charge lattice fluctuations [16] indicated by: a) an anomalous coexistence of superconductivity with short range polaronic Charge Density Waves (CDW) [17,18] and b) an anomalous doping dependent isotope effect [19-24] while in standard BCS superconductors the isotope coefficient is predicted to be independent on the chemical potential and on pressure. The majority of the scientific community considers high temperature superconducting phases in an effective single band in the dirty limit where the controversy since 1987 is between first theories proposing a pairing mechanism mediated by spin fluctuations against other theories proposing a pairing mediated by lattice fluctuations. An alternative third heretic school proposed an alternative mechanism for high temperature superconductivity in a complex phase [16,17] based on the scenario of a superconducting phase in the clean limit made of coexisting condensates of polarons [25] and free carriers [26]. The scenario of superconductivity in the clean limit was confirmed experimentally by the discovery of multigap superconductivity with $T_c$ of 39K in $MgB_2$ [27-30]. Here experiments clearly have shown the coexistence of a narrow σ band in the strong coupling regime and a wide π band in the very weak coupling regime [29-31]. In this scenario a new key term contributes to the condensation energy: the interband interaction between the two condensates [26-33]. From the above the obvious question arises: Can superconductivity in pressurized $H_2S$ at 203 K take place in a single effective band in the dirty limit described by conventional Eliashberg superconductivity made of a single condensate [34-36] or do we need a multi-gap complex scenario in the clean limit ?

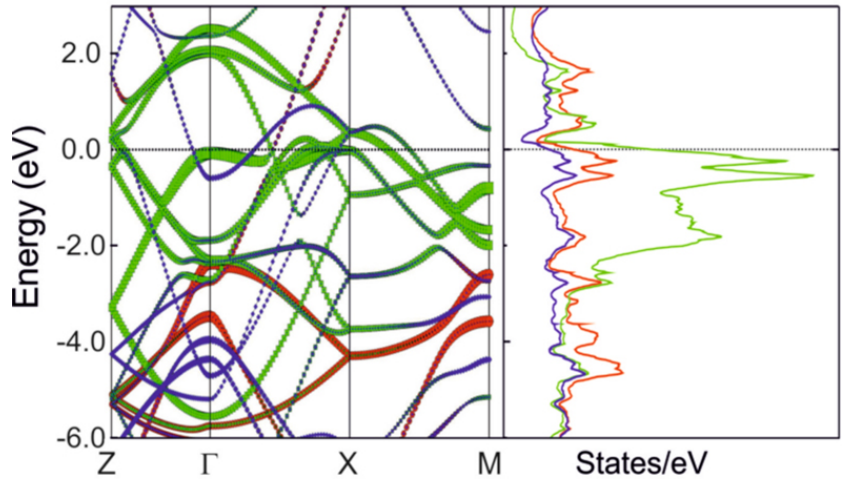

**Figure 2** (Left panel) The electronic band structure of cubic $H_2S$ [37] with the perovskite structure $(SH)^-(H_3S)^+$. (Right panel) The density of states corresponding to the left panel.

Natural limitations of $T_c$ are indicated within BCS theory using the Allen-Dynes equation [36] and Eliashberg [37] theory both of which are based on a pairing mechanism mediated by phonons. Given the inherent relation between phonon frequency ω and electron phonon coupling constant λ as ω≈1/λ it becomes immediately evident that large values of ω imply small values of λ and vice versa. The calculated $T_c$ as a function of λ is presented in Figure 1 for both approaches. While the Allen-Dynes formula yields an increase (though reducing with increasing λ) in $T_c$ with increasing λ, Eliashberg theory finds a maximum $T_c$ around λ=0.9 followed by a decrease with increasing electron-phonon coupling. It is, however, obvious from both results that it is impossible to reach values of $T_c$ above 200 K unless multigap aspects are included as proposed long ago [26-33].

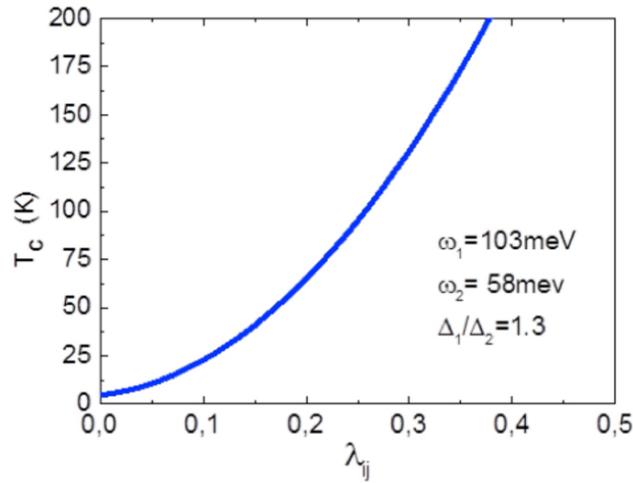

**Figure 3** Superconducting transition temperature $T_c$ as a function of the interband electron-phonon coupling parameter $\lambda_{ij}$ using the cutoff frequency and gap relations as indicated in the figure.

## 2. Multi gap superconductivity in the flat band / steep band scenario

Recently further experimental data on the evolution of the critical temperature of pressurized sulfur hydride as a function of pressure have been reported [37] and it has been possible to extract the pressure dependence of the isotope coefficient [38-40] ranging between 1.5 and 0.2 showing the breakdown of the standard BCS approximations which requires a description of the superconducting phase made of multiple electronic components [41]. In the recent communications Jarlborg et al. [38-40] have calculated the band structure of pressurized sulfur hydride from which they conclude that a flat band / steep band scenario is realized analogous to the case of MgB$_2$, picnides, chalcogens and cuprates [42-49]. The chemical bonding in chalcogens is characterized by $\sigma$ bonds between the atoms, and the $s^2p^4$ configuration splits into low-lying core-like $s^2$ states and p-type valence states. Detailed studies on tellurium [44] show that $p^2$-type lone pairs in the semiconductor are weakly stabilized by approximately 1eV, and become itinerant in the phase under pressure above 4 GPa. The metallic phases exhibit superconductivity up to 7.5K (the lighter homologue sulfur reaches 17K). Electronic band structure calculations reveal a typical flat-band/steep-band scenario [44,45]. The pronounced feature of this scenario in the case of pressurized H$_2$S (see Figure 2) suggests that the hydrogen as well as the sulfur atoms play a crucial role in the high temperature superconducting phase. The band structure of pressurized H$_2$S for its recently presented new structure type, namely the cubic perovskite SH$^-$ and SH$_3^+$is shown in Fig.2 [46]. The fluid like motion of the protons allows for multiple degenerate structural configurations. For one of those the results in Figure 2 have been obtained.

The flat band / steep band scenario requires the use of a minimally two-band approach for superconductivity in pressurized $H_2S$ with typical Hamiltonian [32,33]:

$$H = \sum_k \varepsilon_i(k) c_{ik}^+ c_{ik} + \sum_k \varepsilon_j(k) c_{jk}^+ c_{jk} - \sum_{k,k'} V_{ii} c_{i,k\uparrow}^+ c_{i,-k\downarrow}^+ c_{i,-k'\downarrow} c_{i,k'\uparrow}$$
$$- \sum_{k,k'} V_{jj} c_{j,k\uparrow}^+ c_{j,-k\downarrow}^+ c_{j,-k'\downarrow} c_{j,k'\uparrow} - \sum_{k,k'} V_{ij} c_{i,k\uparrow}^+ c_{i,-k\downarrow}^+ c_{j,-k'\downarrow} c_{j,k'\uparrow}$$
$$- \sum_{k,k'} V_{ji} c_{j,k\uparrow}^+ c_{j,-k\downarrow}^+ c_{i,-k'\downarrow} c_{i,k'\uparrow} \tag{1}$$

Here $\varepsilon$ is the momentum $k$ and band $i$ dependent site energy, $V_{ii}$ and $V_{ij}$ are effectively attractive intra- and interband interactions and $c^+$, $c$ are creation and annihilation operators. The superconducting gap $\Delta_i$ ($i = 1,2$) equations are derived as:

$$\Delta_i(k) = \lambda_{ii} \int_0^{\hbar\omega_i} \frac{\Delta_i(k')}{2E_i(k')} d\varepsilon_i \tanh\left(\frac{E_i(k')}{2k_B T}\right) + \lambda_{ij} \int_0^{\hbar\omega_j} \frac{\Delta_j(k')}{2E_j(k')} d\varepsilon_j \tanh\left(\frac{E_j(k')}{2k_B T}\right) \tag{2}$$

where $E_i = \sqrt{\varepsilon_i^2 + \Delta_i^2}$ and $\omega_i$ is the cutoff frequency of band $i$. These have to be solved self-consistently and simultaneously for each temperature $T$. An approximate analytical expression for $T_c$ can be derived which is given by:

$$kT_c = \sqrt{C\omega_i\omega_j} exp\left[-\frac{1}{2}\frac{\lambda_{ii}+\lambda_{jj}}{\tilde{\lambda}} + \left(\frac{1}{4}\left\{ln\omega_i + ln\omega_j + \frac{\lambda_{ii}+\lambda_{jj}}{\tilde{\lambda}}\right\}^2 - \frac{1}{\tilde{\lambda}}\right)^{1/2}\right] \tag{3}$$

where $\tilde{\lambda} = \lambda_{ii}\lambda_{jj} - \lambda_{ij}\lambda_{ji}$ is a combination of interband and intraband interaction parameters. In order to highlight the role of the interband coupling, $T_c$ is shown as a function of $\lambda_{ij} = \lambda_{ji} = V_{ij}N(E_F)$ with $N(E_F)$ the density of states at the Fermi energy and the intraband couplings being small and fixed in Figure 3. We consider a case where for zero interband coupling $T_c$ is 5 K and rapidly increases to reach 200 K for a value of the interband coupling $\lambda_{ij}$ as small as 0.38 evidencing its important role for achieving high temperature superconductivity. The explicit dependence of $T_c$ on the phonon frequencies in equation 3 allows to calculate the isotope effect on $T_c$ within this approach. It can be obtained in two different ways, namely by varying $\omega_i$ or the phonon mediated interband interaction. Both approaches are shown in Figure 4, where 4a depicts the first case, whereas 4b refers to the latter one.

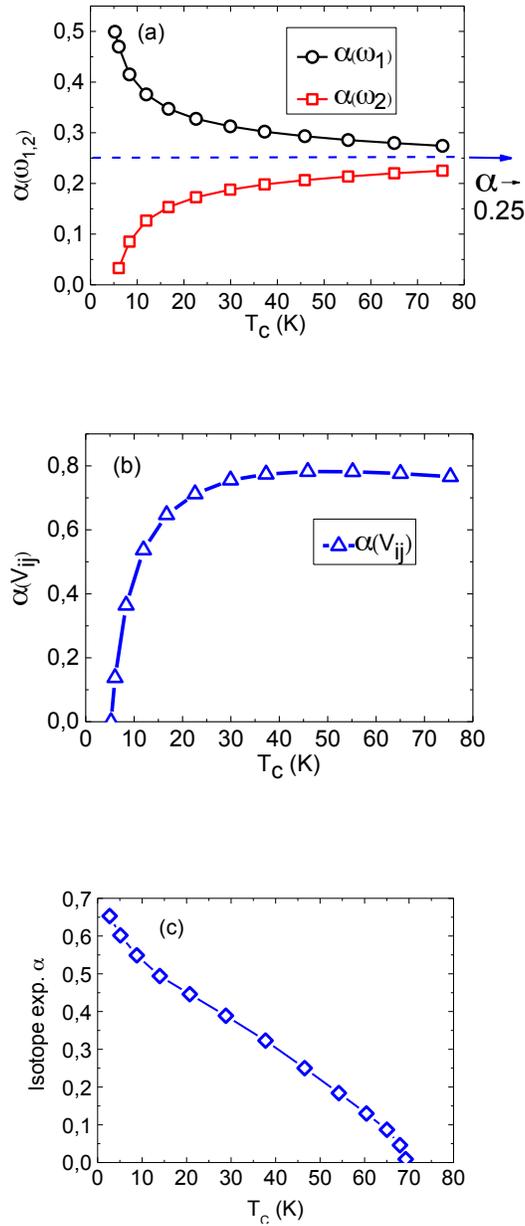

**Figure 4 panel a**): The isotope coefficient $\alpha$ as a function of $T_c$ derived for the intraband phonon contributions: **panel b**): The isotope exponent $\alpha$ as derived for the interband phonon contribution; **panel c**): Isotope exponent $\alpha$ as a function of $T_c/T_{c,\,max}$ as obtained from a two-band polaronic model

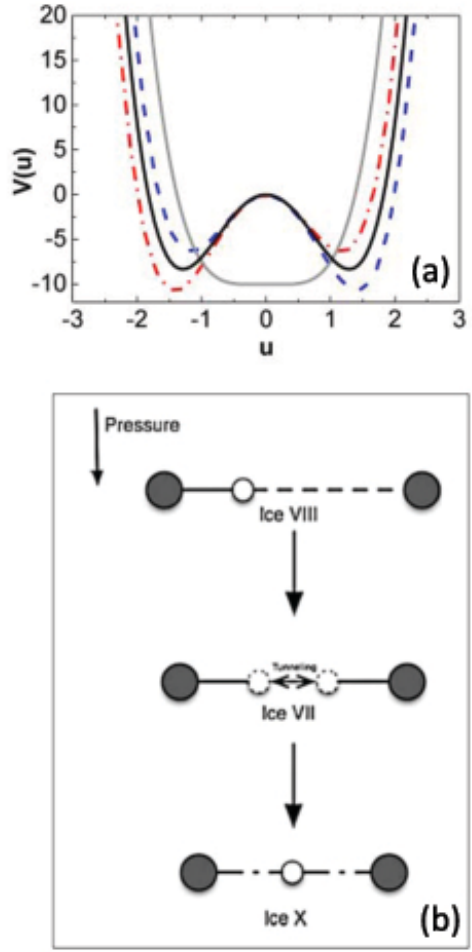

**Figure 5** (a) Schematic changes of the hydrogen related potential with pressure: red and blue at ambient pressure, black at intermediate pressure and grey at high pressure. (b) Example of pressure dependence of the hydrogen atom location in a double well in ice (increasing pressure from above to below).

The variation of the intraband related phonon frequencies has a maximum isotope effect of the BCS value, 0.5, if one of the intraband couplings is zero as expected. For both couplings being finite together with high values of $T_c$ a contribution of 0.25 from each band is achieved whereas the sum of both exponents is always smaller than 0.5. On the other hand $\alpha(\lambda_{ij})$ is small for $T_c$ approaching the single band limit to increase fast with increasing $T_c$ and exceeding the BCS value already for $T_c \approx 15K$. Another approach to calculate the isotope exponent stems from a polaronic renormalization of the band energies, analogous to the case of cuprates. In this case the kinetic energy is reduced exponentially and isotope dependent through the mass renormalization. By considering a constant polaronic coupling and varying $T_c$ through the interband coupling $\alpha$ as a function of $T_c/T_{c,\,max}$ is evaluated and shown in Figure 4c. While the results shown in Figure 4a, 4b do not follow the experimentally observed pressure dependent isotope effect in pressurized $H_2S$ [38], the polaronic approach is in qualitative agreement with those. Relating these results with experimental ones for $H_2S$ under high pressure they can be correlated with the relevance of the interband coupling between narrow and wide bands [26-33, 38-49]

An important ingredient in the above described multi-band approach to superconductivity is the involvement of multi phonons [50,51], since the pairing potentials within the bands do not arise from the same phonons. In addition, upon performing a Bogoliubov transformation of equation 1 with the assumption that the pairing is lattice mediated, reveals that a single phonon process cannot account for the interband pairing. However, multi-phonon processes are a typical signature for strong lattice anharmonicity as realized, e.g., in ferroelectrics [52]. For pressurized $H_2S$ anharmonicity in the lattice dynamics has been assigned as arising from the hydrogen bonds [38,42]. Indeed, the hydrogen bond typically exhibits a double-well potential which is asymmetrical and randomly occupied at elevated temperature or ambient pressure [51]. With decreasing temperature/increasing pressure it adopts a symmetrical shape which merges to a single well potential at low temperature/high pressure (Figure 5a). This is exemplified for the case of $H_2O$ under pressure in Figure 5b. It implies that any kind of anharmonicity of the hydrogen bonds vanishes in $H_2S$ in the high pressure range. Nevertheless anharmonicity remains important, however, for the sulfur atom which bridges the two hydrogen atoms and strongly hybridizes its p-orbitals with the s-orbitals of H. This hybridization induces a double-well potential in the sulfur motion where a pseudo-harmonic approximation can be used to study its influence on superconductivity. This results in the following regimes for the involved phonon frequencies:

$$\omega^2 = \frac{k}{m} : harmonic\ limit \qquad (4a)$$

$$\omega^2 = \frac{k}{m} + \frac{k_4}{m} \langle u^2 \rangle : anharmonic\ limit \qquad (4b)$$

where

$$\langle u^2 \rangle_T = \sum_{q,j} \frac{\hbar}{\omega(q,j)} U^2(q,j) coth \frac{\hbar\omega(q,j)}{2kT} \qquad (4c)$$

The consequences for superconductivity are readily obtained since in equ. 3 either of the phonon mode frequencies is modified by the nonlinearity and given by:

$$\omega_i^2 = \frac{k}{m} + \frac{k_4}{m} \langle u^2 \rangle \qquad (5)$$

Besides the role of the interband interaction the nonlinearity in the involved phonon modes thus also contributes to a substantial enhancement of $T_c$ together with important variations of the isotope effect with pressure.

A. Bussmann-Holder, J. Köhler, M.-H. Whangbo, A. Bianconi, A. Simon.

## 3. Conclusions

In this work we argue that the high value of $T_c$ in pressurized sulfur hydride is incompatible with conventional BCS or Eliashberg type pairing interactions. While conventional BCS theory predicts a pressure independent isotope coefficient the experimental dependence of the isotope effect on pressure [38-40] supports the breakdown of standard BCS approximations in describing the superconductivity in pressurized sulfur hydride. We suggest a complex superconductivity landscape made of multiple condensates which can explain the $T_c$'s as large as 200K. Anharmonicity supports pressure dependent isotope effects and $T_c$ enhancements. Deviations from BCS predictions concerning coherence factors, as they are apparent in multiband superconductors, are a useful test to confirm multigap superconductivity. Further work is needed to include in the complex universe of multiple condensates the spatial topology due to the arrested phase separation near Lifshitz transitions for strongly interacting electronic systems [18,53-57].